\documentclass[conference,10pt]{IEEEtran}
\IEEEoverridecommandlockouts
\usepackage{cite}
\usepackage{amsmath,amssymb,amsfonts}
\usepackage{algorithmic}
\usepackage{graphicx,physics,array}
\usepackage{textcomp}
\usepackage{xcolor,soul}
\usepackage{multirow}
\usepackage{hyperref}
\usepackage{adjustbox}
\usepackage{tikz}

\newcommand*\whitecircled[1]{\tikz[baseline=(char.base)]{
            \node[shape=circle,draw,fill=white,inner sep=0.5pt] (char) {\textcolor{black}{#1}};}}

\def\BibTeX{{\rm B\kern-.05em{\sc i\kern-.025em b}\kern-.08em
    T\kern-.1667em\lower.7ex\hbox{E}\kern-.125emX}}

\begin{document}

\title{Quantum Prometheus: Defying Overhead with Recycled Ancillas in Quantum Error Correction
}

\author{\IEEEauthorblockN{Avimita Chatterjee}
\IEEEauthorblockA{\textit{Department of CSE} \\
\textit{Pennsylvania State University}\\
PA, USA\\
amc8313@psu.edu}
\and
\IEEEauthorblockN{Archisman Ghosh}
\IEEEauthorblockA{\textit{Department of CSE} \\
\textit{Pennsylvania State University}\\
PA, USA\\
apg6127@psu.edu}
\and
\IEEEauthorblockN{Swaroop Ghosh}
\IEEEauthorblockA{\textit{School of EECS} \\
\textit{Pennsylvania State University}\\
PA, USA\\
szg212@psu.edu}
}

\maketitle

\begin{abstract}
Quantum error correction (QEC) is crucial for ensuring the reliability of quantum computers. However, implementing QEC often requires a significant number of qubits, leading to substantial overhead. One of the major challenges in quantum computing is reducing this overhead, especially since QEC codes depend heavily on ancilla qubits for stabilizer measurements. In this work, we propose reducing the number of ancilla qubits by reusing the same ancilla qubits for both X- and Z-type stabilizers. This is achieved by alternating between X and Z stabilizer measurements during each half-round, cutting the number of required ancilla qubits in half. This technique can be applied broadly across various QEC codes, we focus on rotated surface codes only and achieve nearly \(25\%\) reduction in total qubit overhead. We also present a few use cases where the proposed idea enables the usage of higher-distance surface codes at a relatively lesser qubit count. Our analysis shows that the modified approach enables users to achieve similar or better error correction with fewer qubits, especially for higher distances (\(d \geq 13\)). Additionally, we identify conditions where the modified code allows for extended distances (\(d + k\)) while using the same or fewer resources as the original, offering a scalable and practical solution for quantum error correction. These findings emphasize the modified surface code's potential to optimize qubit usage in resource-constrained quantum systems.


\end{abstract}

\begin{IEEEkeywords}
Quantum error correction codes, surface codes, distance, physical error rate, logical error rate, ancilla qubit, data qubit, threshold. 
\end{IEEEkeywords}

\section{Introduction}

Quantum Error Correction (QEC) is a fundamental requirement for the fault-tolerant operation of quantum computers. In the delicate environment of quantum computation, where information is encoded in fragile quantum states, even the slightest interaction with the surrounding environment can induce errors that compromise computational accuracy~\cite{mouloudakis2021entanglement}. Quantum systems are particularly vulnerable to two primary types of errors: bit-flip and phase-flip errors. Unlike classical systems, which primarily contend with bit-flip errors, quantum systems face the additional complexity of phase-flip errors due to the unique properties of quantum information. Traditional error correction techniques, as applied in classical systems, are ineffective for quantum data because of the constraints imposed by the no-cloning theorem and the disruptive nature of quantum measurement~\cite{wootters2009no, von2018mathematical}. This intrinsic vulnerability of quantum states underscores the necessity for specialized QEC techniques tailored to the quantum realm.

To counteract these errors, QEC introduces various stabilizer codes, which employ ancillary, or ``ancilla," qubits to measure and correct errors without disturbing the encoded quantum information. However, the robustness of QEC comes at a significant cost: the need for numerous ancilla qubits, which imposes substantial overhead on quantum systems and presents a major barrier to scalable quantum computation. Fig.~\ref{fig:everything_basic} \whitecircled{c} illustrates the increase in the total number of qubits and data qubits as the distance of rotated surface codes grows. The total number of qubits is significantly higher than the number of data qubits, which can be attributed to the ancilla qubits. For a given distance, the number of ancilla qubits is approximately equal to the number of data qubits.

This paper introduces an \emph{efficient approach to counter the overhead of numerous ancilla qubits by reusing the ancilla qubits for stabilizer measurements}. Since surface codes~\cite{dennis2002topological} address both bit and phase errors by using a two-dimensional qubit layout, it makes them a frontrunner for near-term fault-tolerant quantum computing.  Within this category, there are `rotated'~\cite{tomita2014low} and `unrotated'~\cite{fowler2012surface} variants. Rotated surface codes offer a higher resistance to errors and are somewhat less complex to implement than the unrotated versions thus, we focus exclusively on rotated surface codes in this work. Since the basic anatomy of a quantum error correction code (QECC) involves rounds and stabilizer circuits with ancilla qubits, the proposed approach can be extended to other QECCs seamlessly.

The paper is organized as follows: Section~\ref{sec:background} provides an overview of surface codes and the motivation for the proposed approach. Section~\ref{sec:modified_surface_code} presents the methodology for the modified surface code. Section~\ref{sec:use_case_math} explores illustrative use cases, offering a detailed mathematical generalization of qubit usage for various scenarios. Section~\ref{sec:evaluation} compares the performance of the modified and original surface codes. Finally, Section~\ref{sec:conclusion} concludes the paper.
\section{Background and Motivation} \label{sec:background}

\begin{figure*}
    \centering
    \includegraphics[width=1\linewidth]{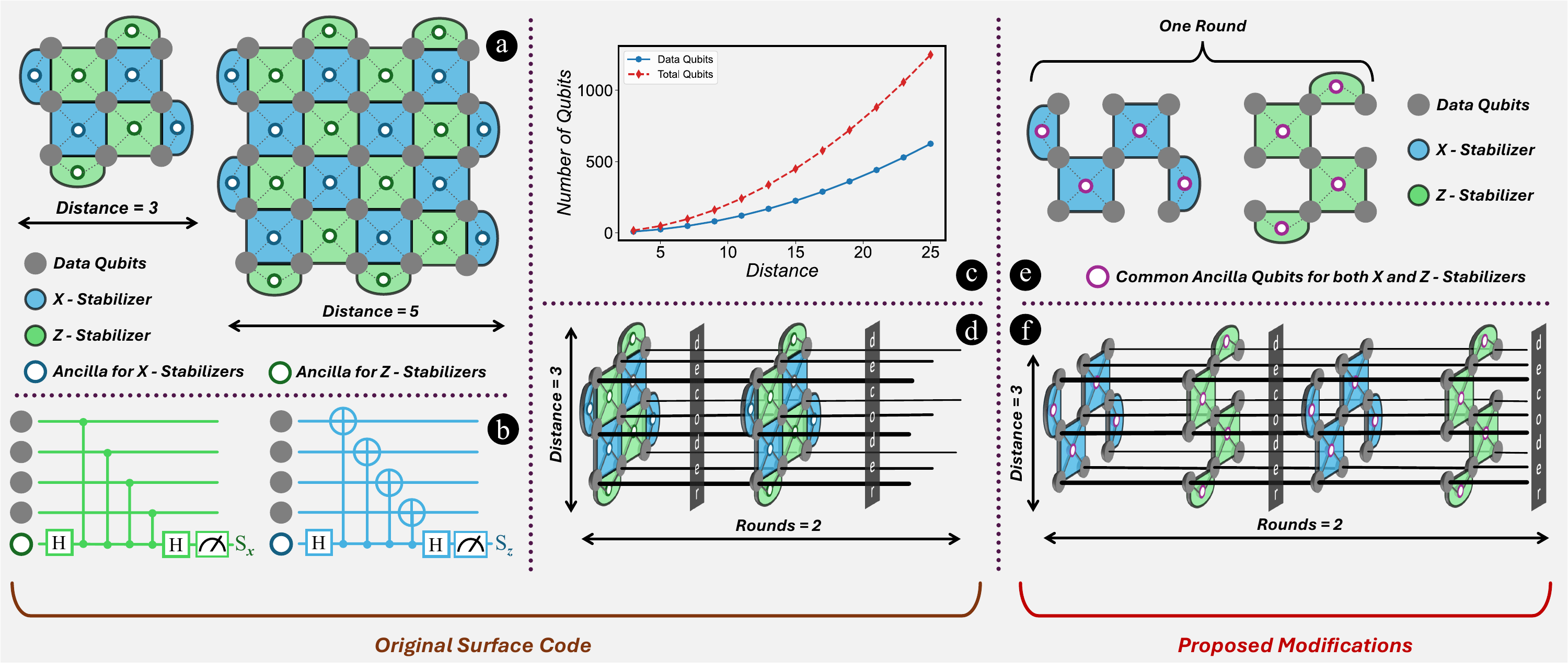}
    \caption{\textbf{Original Approaches in the Rotated Surface Code and the Proposed Modification Strategy}
     (a) A distance-3 and distance-5 rotated surface code layout, demonstrating the scalable structure of the code that enhances fault tolerance with increased distance. The layout shows how expanding the code distance, by adding additional qubits, increases robustness against errors.
     (b) Example of Z- and X-stabilizer circuits with ancilla measurements responsible for detecting bit-flip and phase-flip errors respectively.
     (c) Growth in qubit requirements with increasing code distance, illustrating the substantial overhead introduced by ancilla qubits. As the distance of the code increases to improve error tolerance, the number of qubits required for both data and ancilla qubits rises, with ancilla qubits contributing significantly to the total overhead.
     (d) Distance-3 rotated surface code shown over two rounds with decoding cycles.
     (e) Proposed approach of dividing each round into two sub-rounds, where ancilla qubits are repurposed. In this approach, the first sub-round measures only X-stabilizers, while the second sub-round measures only Z-stabilizers, using the same set of ancilla qubits. This strategy reduces the required number of ancilla qubits by half, effectively lowering the qubit overhead.
     (f) Modified surface code operation over two rounds, showing each round containing two sub-rounds and only one decoder after every complete round.
    }
    \label{fig:everything_basic}
\end{figure*}

The structure of most quantum error-correcting codes (QECCs) is anchored in two main stabilizers: the X-stabilizer and the Z-stabilizer. X-stabilizers detect phase-flip (Z-flip) errors, while Z-stabilizers identify bit-flip (X-flip) errors. In Fig.~\ref{fig:everything_basic} \whitecircled{a}, a surface code with distance 3 is shown, illustrating how this layout works. Each ``distance" represents the code's error-tolerance level, with larger distances enhancing fault tolerance; this setup can scale to higher distances, such as distance 5, to increase robustness against errors. The surface code is arranged on a $d \times d$ lattice, where each gray blob represents a data qubit. Every data qubit is monitored by both X and Z stabilizers, ensuring complete coverage for detecting both bit-flip and phase-flip errors. The green regions represent Z-stabilizers, while the blue regions indicate X-stabilizers. Ancilla qubits, used to capture error syndromes, are depicted as white blobs with borders: green-bordered white blobs for Z-stabilizers and blue-bordered white blobs for X-stabilizers.

In this arrangement, each stabilizer interacts with a set of four neighboring data qubits to generate a syndrome as shown in Fig.~\ref{fig:everything_basic} \whitecircled{b}. For example, a Z-stabilizer produces the syndrome $Z(q_0) \otimes Z(q_1) \otimes Z(q_2) \otimes Z(q_3)$, which is then projected onto an ancilla qubit. This projection yields a measurement $S_z$, indicating a $\pm 1$ outcome depending on the presence of a bit-flip error in the associated data qubits. Similarly, the X-stabilizer produces the syndrome $X(q_0) \otimes X(q_1) \otimes X(q_2) \otimes X(q_3)$ with the ancilla measurement $S_x$ indicating any phase-flip error. 

In this configuration, each stabilizer in the surface code has a corresponding ancilla qubit to capture its syndrome measurement—half of these stabilizers are designated as X-stabilizers, and the other half as Z-stabilizers. In a rotated surface code of distance \( d \), there are \( d^2 \) data qubits arranged in a \( d \times d \) grid. The stabilizers, which require ancilla qubits for syndrome measurement, are positioned both within the grid and along the boundaries. The number of squares (or cells) inside the grid that host stabilizers is \( (d - 1)^2 \), with each square containing either an X-stabilizer or a Z-stabilizer. Additionally, there are boundary stabilizers: \( d - 1 \) semi-circular X-stabilizers along the left and right edges and \( d - 1 \) semi-circular Z-stabilizers along the top and bottom edges. Summing these, the total number of stabilizers is \( (d - 1)^2 + 2(d - 1) = d^2 - 1 \). Therefore, \( d^2 - 1 \) ancilla qubits are needed to monitor these stabilizers. Combining the data and ancilla qubits, the total number of qubits in a rotated surface code of distance \( d \) is \( 2d^2 - 1 \). This result highlights the substantial resource demand from ancilla qubits in quantum error correction, as they are nearly equal in number to the data qubits, making efficient management of qubit overhead critical in scalable quantum computing. The total number of qubits required for a rotated surface code of distance \( d \) is given by: \( Q_{\text{total}} = 2d^2 - 1 \). Fig.~\ref{fig:everything_basic} \whitecircled{c} demonstrates the increase in data qubits as the distance of a rotated surface code is expanded. However, the overall rise in the total number of qubits is substantially greater, as it includes both data and ancilla qubits. Therefore it is important to explore strategies to reduce the overall qubit overhead to make the quantum error correction scheme scalable.
Although X and Z stabilizers are measured simultaneously, the ancilla qubits used for each are separate. This means that while X-stabilizers are being measured, the Z-stabilizers utilize a completely different set of ancilla qubits. This prompts us to design a stabilizer circuit with \emph{common} ancilla qubits to measure both the X and Z-stabilizers thus reducing the qubit overhead and making the surface code more scalable.

In a traditional implementation of the surface code, a single application of the code is insufficient for effective error correction. Instead, the surface code is applied repeatedly in sequential steps, known as rounds, on the same set of qubits. After each round, a decoding algorithm is executed to detect and correct errors based on the syndromes measured in that round. Once corrections are made, the process proceeds to the next round. This cycle can continue for as many rounds as needed to ensure stability and accuracy in the quantum computation. Fig.~\ref{fig:everything_basic} \whitecircled{d} shows a distance-3 rotated surface code running for two rounds, with decoding cycles at the end of each round.
\section{Modified Surface Code and Use Cases} \label{sec:modified_surface_code}

In this section, we introduce a modified surface code to significantly reduce the number of ancilla qubits without compromising the code's structural integrity or error-detection capabilities. We also present various use cases where the proposed approach may be attractive.
\subsection{Proposed Approach}
Traditionally, a surface code round consists of a simultaneous measurement of both X-stabilizers and Z-stabilizers, each requiring a dedicated set of ancilla qubits. However, as depicted in Fig.~\ref{fig:everything_basic} \whitecircled{e}, we divide each standard round into two sequential sub-rounds. In the first sub-round, only X-stabilizers are measured, utilizing a set of ancilla qubits. In the second sub-round, the same ancilla qubits are repurposed to measure Z-stabilizers. This reuse of ancilla qubits, marked by the purple circles in the figure, allows for a reduction in the number of ancilla qubits by half after performing reset operations on the ancilla qubits before reuse.
Fig.~\ref{fig:everything_basic} \whitecircled{f} illustrates the above approach for two full rounds. 


\begin{figure}
    \centering
    \includegraphics[width=1\linewidth]{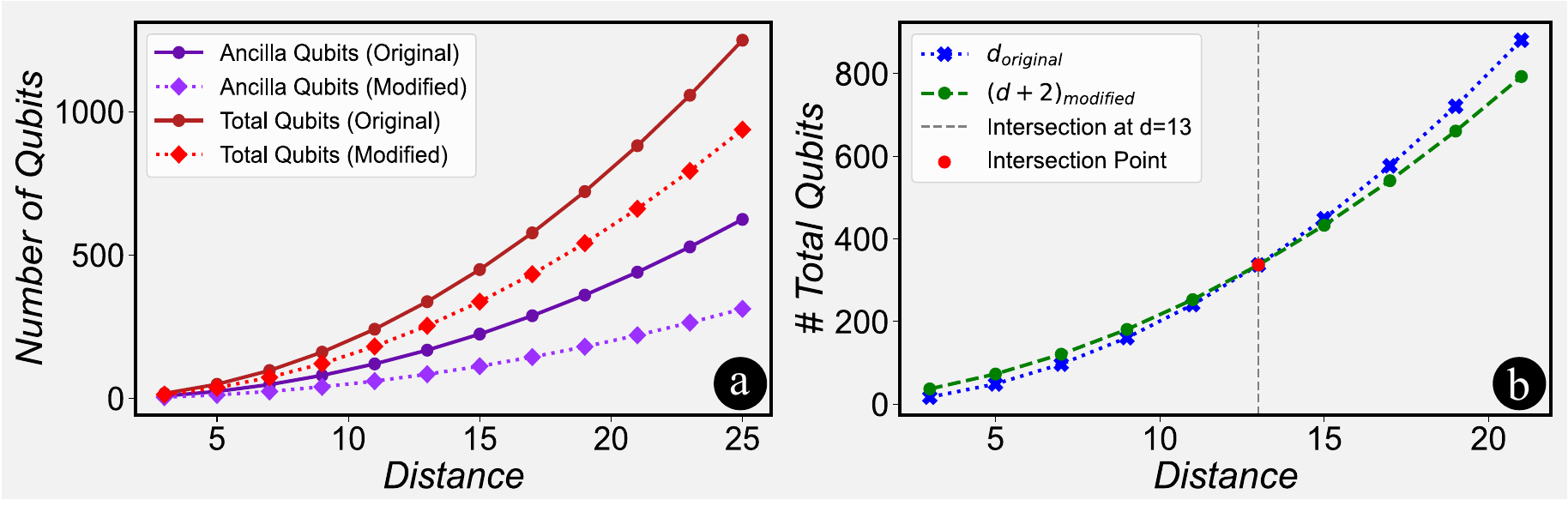}
    \caption{\textbf{Comparison of Qubit Overhead and Ratios in Original and Modified Surface Code Approaches}
    (a) compares ancilla and total qubit requirements in the original and new approaches, showing that while both increase with code distance, the new approach achieves a marked reduction in ancilla qubits, leading to a lower total qubit count. This demonstrates the effectiveness of recycling ancillas in reducing overhead, particularly at higher distances.
    (b) illustrates the total number of qubits as a function of the surface code distance, comparing \(d\) (original) and \(d + 2\) (modified). Notably, the lines intersect at \(d = 13\), indicating the point where the total qubits required for \(d + 2\) (modified) become equal to those for \(d\) (original). Beyond this intersection, for \(d \geq 13\), the modified approach requires fewer qubits than the original, highlighting its resource efficiency at higher distances.
    }
    \label{fig:new_overhead_and_ratios}
\vspace{-13
    pt}
\end{figure}
In the traditional surface code configuration, each stabilizer (both X and Z) requires a separate ancilla qubit for syndrome measurement, resulting in a total of \( d^2 - 1 \) ancilla qubits for a code of distance \( d \). 
However, 
our modified surface code requires only approximately \( \frac{d^2 - 1}{2} \) ancilla qubits. Thus, the reduction in the number of ancilla qubits \( R_{\text{ancilla}} \) is 50\%, which lowers the overall resource demands of the code by $\sim25\%$ while preserving its error-correcting capabilities. 
Fig.~\ref{fig:new_overhead_and_ratios} \whitecircled{a} compares four curves: ancilla qubits (original and the modified approach) and total qubits (original and modified). Both ancilla and total qubits increase with distance, but the modified approach shows a marked reduction in ancilla qubits, leading to a corresponding decrease in the total qubit count. 

\subsection{Illustrative Use Cases} \label{sec:use_case_math}

In practical applications of quantum error correction, users often face trade-offs between achieving a specific logical error rate and minimizing qubit overhead. The modified surface code offers a way to reduce the number of ancilla qubits required, making it a compelling choice in certain scenarios.

\subsubsection{\textbf{Case A: Relaxed Logical Error Rate Requirements}}

Suppose a user needs to use a distance-3 surface code to achieve a target logical error rate. If the user can tolerate a slightly higher logical error rate than planned but lacks the required number of qubits, they can switch to the modified method. This approach enables the user to employ distance-d surface code with their available resources while impacting approximately a 25\% reduction in total qubit overhead while still maintaining reasonable error-correcting capabilities.

\subsubsection{\textbf{Case B: Strict Logical Error Rate Requirements}}

In scenarios where the user aims to achieve the logical error rate associated with a surface code of distance \(d\) (original), using a modified surface code of the same distance will result in a slightly higher logical error rate. To meet the logical error rate requirement while reducing qubit overhead, the user can transition to a modified surface code of distance \(d + 2\).

The additional qubits required for this transition are calculated as:
\( \Delta Q = Q_{\text{modified}, d+2} - Q_{\text{original}, d}\).
Substituting the formulas for \(Q_{\text{modified}}\) and \(Q_{\text{original}}\), we obtain:
\[
\Delta Q = \left[ (d + 2)^2 + \frac{(d + 2)^2 - 1}{2} \right] - \left[ d^2 + (d^2 - 1) \right]
\]
\[
\Delta Q = \frac{1}{2}(d + 1)(13 - d)
\]

\paragraph{For distances \(d < 13\)}
For distances \(d < 13\), this equation shows a \textit{positive value of \(\Delta Q\)}, meaning the user will need \textit{additional qubits} to transition from \(d\) (original) to \(d + 2\) (modified). However, this additional qubit requirement is less than what would have been needed to directly implement a surface code of distance \(d + 2\) (original). For instance, if \(d = 3\): \(\Delta Q = 37 - 17 = 20\).
In this example, the user needs 20 additional qubits to transition from \(d = 3\) (original) to \(d = 5\) (modified), achieving the logical error rate of a distance-5 code. This approach saves resources compared to directly implementing a distance-5 code (original), which would require 32 qubits instead of 37.

\paragraph{For distances \(d \geq 13\)}

When the distance \(d\) (original) is large (\(d \geq 13\)), the situation changes significantly. 
If \(d \geq 13\), the value of \(\Delta Q\) becomes \textit{negative}, indicating that transitioning from \(d\) (original) to \(d + 2\) (modified) requires \textit{fewer qubits}. This happens because the qubit-saving efficiency of the modified code at larger distances outweighs the additional resources needed for the increase in distance (as shown in Fig.~\ref{fig:new_overhead_and_ratios} \whitecircled{b}).

For example, if \(d = 13\):
\(
\Delta Q = 337 - 337 = 0
\)
At \(d = 13\), the qubit requirements for \(d\) (original) and \(d + 2\) (modified) are identical.
For \(d > 13\), the modified code of \(d + 2\) requires \textit{fewer qubits} than \(d\) (original). For instance, if \(d = 15\):
\(
\Delta Q = 433 - 449 = -16
\)
In this case, transitioning from \(d = 15\) (original) to \(d = 17\) (modified) results in saving 16 qubits. The modified approach becomes even more efficient, enabling the user to achieve a similar or lower logical error rate of \(d + 2\) (modified) with \textit{fewer qubits} than required for \(d\) (original). This makes the modified approach particularly advantageous. 

\subsubsection{\textbf{Case C: Maximizing Gains with Fixed Qubit Resources}}

\begin{table}[]
\centering
\caption{Comparison of Qubit Usage and Logical Error Rates for \(d\) (Original) and \(D = d+k\) (Modified) Schemes at a Physical Error Rate of \(10^{-3}\)}
\begin{tabular}{ccc||c||cccc}
\multicolumn{3}{c||}{\textbf{Original}}        & \multirow{2}{*}{\textbf{k}} & \multicolumn{4}{c}{\textbf{Modified Approach}}                             \\ \cline{1-3} \cline{5-8} 
\textbf{d}  & \textbf{\#Q} & \textbf{Logical} &                             & \textbf{D}  & \textbf{\#Q} & \textbf{Logical} & \textbf{\# Saved} \\ \hline \hline
\textbf{13} & 337          & 9.52e-13         & 2                           & \textbf{15} & 337          & 1.95e-14         & 0                 \\
\textbf{27} & 1457         & 4.27e-24         & 4                           & \textbf{31} & 1441         & 1.50e-27         & 16                \\
\textbf{39} & 3041         & 8.01e-34         & 6                           & \textbf{45} & 3037         & 5.06e-39         & 4                 \\
\textbf{53} & 5617         & 3.55e-45         & 8                           & \textbf{61} & 5581         & 3.91e-52         & 36                \\ \hline \hline
\end{tabular}
\label{tab:determin_k_for_modified}
\end{table}

In this scenario, we determine how far a user can increase the distance \(k\) in a modified surface code (\(d + k\)) while using the same or fewer qubits than required for a distance \(d\) (original) surface code. This analysis highlights the flexibility of the modified surface code in leveraging additional error-correcting performance with fewer resources.

The total qubit count for a modified surface code of distance \(d + k\) is:
\(
Q_{\text{modified}, d+k} = (d + k)^2 + \frac{(d + k)^2 - 1}{2}
\); and the total qubit count for an original surface code of distance \(d\) is:
\(
Q_{\text{original}, d} = d^2 + (d^2 - 1)
\)
To find the maximum \(k\) for which the modified surface code uses the same or fewer qubits as the original, we solve the inequality:
\[
Q_{\text{modified}, d+k} \leq Q_{\text{original}, d}
\]
The maximum \(k\) for which a user can transition to \(d + k\) (modified) while using the same or fewer qubits as \(d\) (original) can be determined by solving the inequality. After simplifying and approximating, we obtain:

\[
k \leq \left(\frac{2}{\sqrt{3}} - 1\right) \cdot d.
\]

We validate the above result by explicitly calculating the difference in total qubits (\(\Delta Q\)) for \(k = 4\) to confirm the transition point.
For \(d = 27\) (original):
\(
Q_{\text{original}, d=27} = 1457
\) and
for \(d + 4 = 31\) (modified):
\(
Q_{\text{modified}, d+4=31} = 1441
\).
Therefore, 
\(
\Delta Q = Q_{\text{modified}, d+4=31} - Q_{\text{original}, d=27} = -16
\)
Since \(\Delta Q < 0\), \(d + 4 = 31\) (modified) uses fewer qubits than \(d = 27\) (original).

Table~\ref{tab:determin_k_for_modified} highlights that the \(d+k\) (modified) scheme uses fewer or equal qubits compared to the \(d\) (original) scheme, while consistently providing a logical error rate lower than that of the \(d\) (Original) scheme. This demonstrates the efficiency of the ``modified" approach in leveraging an increased code distance for better logical error suppression without incurring additional qubit overhead. For instance, when \(d = 27\) in the ``Original" scheme, the logical error rate is \(4.27 \times 10^{-24}\) using 1457 qubits, whereas the ``Modified" scheme with \(d + k = 31\), where \(k = 4\) achieves a lower logical error rate of \(1.50 \times 10^{-27}\) while requiring only 1441 qubits, saving 16 qubits. Importantly, this table assumes a physical error rate of \(10^{-3}\), providing a consistent baseline for comparing the two schemes' performance.
\section{Comparison and Evaluation} \label{sec:evaluation}

In this section, we discuss the experimental setup, the noise modeling approach used for our experiments, and the results, including a comparison of logical error rates, threshold behavior, and performance analysis across different code distances.

\subsection{Experimental Setup}
\begin{figure*}[h]
    \centering
    \includegraphics[width=1\linewidth]{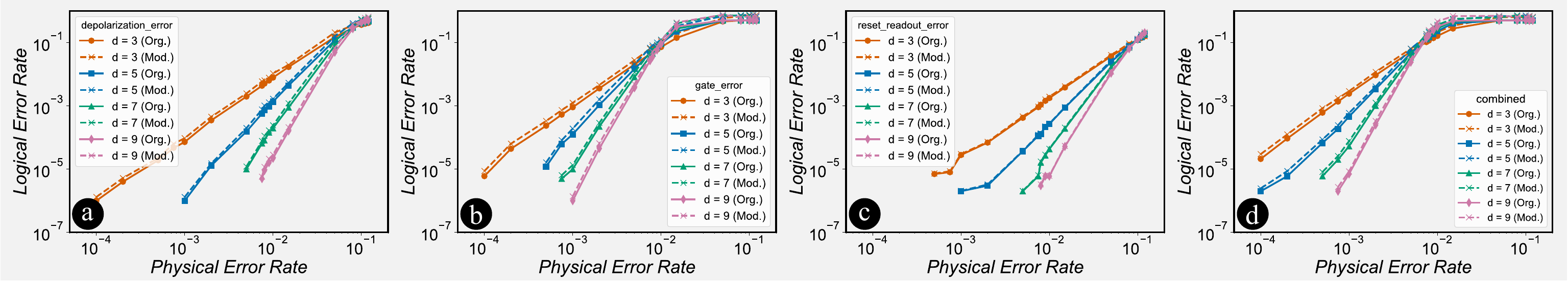}
    \caption{\textbf{Analyzing Logical Error Rates Across Physical Error Rates for Original and Modified Surface Codes}
    To compare the original and modified approaches to surface codes, this figure illustrates the relationship between logical error rate and physical error rate for four distinct types of errors—depolarizing, gate, readout/reset, and combined errors—across original and modified surface codes of distances 3, 5, 7, and 9. 
    (a) shows the impact of depolarizing errors, where higher code distances consistently result in lower logical error rates due to better error correction. This behavior is evident for both original and modified codes, although the modified code exhibits slightly higher logical error rates. A similar pattern is observed for the other error types:
    (b) gate errors, 
    (c) readout/reset errors, and 
    (d) combined errors. 
    In all cases, the modified surface code maintains a logical error rate trend closely resembling the original, demonstrating improved performance with increasing code distance.
    }    \label{fig:logical_vs_physical_all_distance_origiinal_vs_modified}
\end{figure*}
In this study, we utilize STIM \cite{gidney2021stim}, a high-performance simulator for quantum stabilizer circuits, to conduct our experiments. STIM employs a stabilizer tableau representation, akin to the CHP simulator \cite{aaronson2004improved}, while incorporating significant improvements in both performance and accuracy. As highlighted in prior studies \cite{krinner2022realizing, acharya2024quantum}, STIM excels at faithfully simulating the behavior of QECCs on real quantum hardware. For our simulations, we configured STIM to evaluate both original and modified surface codes under various distances, noise types, and physical error rates. The noise models considered include depolarizing, gate, readout, and reset errors, with controlled probabilities to capture a broad spectrum of error scenarios. Logical error rates were extracted from the simulator outputs at sampled physical error rates to illustrate the relationship between logical and physical error rates for each code and noise configuration. 

Given the current limitations on the number of qubits in freely accessible quantum hardware and the incompatibility of available hardware layouts with QECC implementation, direct execution of QECCs on these platforms is not yet feasible. As a result, STIM proves to be an essential tool for advancing research. To ensure consistency with the distance, it is essential to increase the number of rounds proportionally as the distance increases. In our experiments, we used the formula 
\((\text{distance} \times 3) + 1\) 
to determine the number of rounds. This approach ensures that the number of rounds remains even, as the distance is typically an odd number, allowing us to divide each round into two sub-rounds. All logical error rates presented are calculated per round and decoded using the MWPM (Minimum Weight Perfect Matching) algorithm \cite{kolmogorov2009blossom, higgott2022pymatching}.

\subsection{Modeling Noise for Experimental Setup}
\subsubsection{Depolarizing error} A single-qubit depolarizing error picks a random Pauli error ($X$, $Y$, $Z$) based on the coherence time and the quantum gate exposure duration and applied to the qubit. We model depolarizing error to act on all data qubits before every round of applying the surface code as well as after every Clifford gate in the original surface code. To make an even comparison, we apply the depolarizing error before every second sub-round in the modified surface code. This is because two sub-rounds in the modified surface code are equivalent to one round in the original surface code.

\subsubsection{Gate error} Factors like miscalibration and environmental noise cause gate errors to act on the quantum circuit. To model gate error, we apply Pauli-X and Pauli-Z errors on all data qubits after every round of applying the surface code. For single Clifford operations, the execution considers an error probability, and for two-qubit Clifford operations, a random pair of Pauli errors is applied.

\subsubsection{Readout and Reset error} We model readout error as \textit{pre-measurement} probabilistic flip-operations and reset errors are modeled as \textit{post-reset} probabilistic flip-operations to the qubits initialized to the $\ket{0}$ state. They are also applied after the measure, reset, and measure-reset operations.

\subsubsection{Combination} This is simply a combination of all the above-mentioned errors applied to the qubits before applying a surface for error correction. Note that during the combination of the errors the probability of applying each error operation is identical. 

\subsection{Results and Discussion}

\subsubsection{\textbf{Comparison with Respect to Various Error Types}}
In quantum error correction, the physical error rate represents the inherent error present in the system, while the logical error rate denotes the residual error after applying error correction mechanisms. To compare the original and modified approaches to surface codes, it is essential to analyze how the logical error rate varies with the physical error rate. Fig.~\ref{fig:logical_vs_physical_all_distance_origiinal_vs_modified} illustrates this relationship for the four distinct types of errors considered in this work across original and modified surface codes of distances 3, 5, 7, and 9. Subfigure Fig.~\ref{fig:logical_vs_physical_all_distance_origiinal_vs_modified} \whitecircled{a} depicts the effect of depolarizing errors. As expected, higher code distances consistently result in lower logical error rates due to their superior error-correcting capabilities. This trend is observed for both the original and modified surface codes. While the modified code follows a similar trend to the original, it exhibits a slightly higher logical error rate. A comparable pattern is observed for the other error types: Fig.~\ref{fig:logical_vs_physical_all_distance_origiinal_vs_modified} \whitecircled{b} gate errors, Fig.~\ref{fig:logical_vs_physical_all_distance_origiinal_vs_modified} \whitecircled{c} readout/reset errors, and Fig.~\ref{fig:logical_vs_physical_all_distance_origiinal_vs_modified} \whitecircled{d} combined errors. Across all cases, the modified surface code demonstrates a logical error rate trend that closely mirrors the original, maintaining the general behavior of enhanced performance with increased code distance.

While the modified approach reduces the number of required ancilla qubits by half, it introduces temporal correlations between errors in the two rounds. As a result, the modified code, although following a similar trend to the original, exhibits a slightly higher logical error rate. This is because errors accumulated during the \(X\)-stabilizer round can propagate into the \(Z\)-stabilizer round, increasing the likelihood of logical errors. In contrast, the original code avoids this propagation by using separate ancilla qubits for \(X\)- and \(Z\)-stabilizer measurements. The change in logical error rate across error types also varies based on the nature of the errors and their interaction with the reused ancilla qubits. 

To analyze the differences between original and modified logical error rates for various physical error rates across multiple distances and error types, we defined the \textit{relative ratio metric} as:

\[
\text{Relative Ratio} = \frac{\text{Modified Logical Error}}{\text{Original Logical error}}
\]

This metric provides a normalized measure of change, where ratios close to 1 indicate minimal change and ratios greater than or less than 1 indicate proportional increases or decreases, respectively. 
For each combination of error type, distance, and error rate, the relative ratio is calculated to compare the modified and original values. The mean relative ratio for each distance within an error type is then computed across all error rates. To summarize changes for an entire error type, the mean relative ratio is further averaged across all distances.

\begin{table}[]
\centering
\caption{Comparison of Mean Relative Ratios Across Distances for Different Error Types}

\begin{tabular}{c||cccc|c}
\textbf{Error Type} & \textbf{d = 3} & \textbf{d = 5} & \textbf{d = 7} & \textbf{d = 9} & \textbf{Mean}  \\ \hline \hline
Depolarizing        & 1.295          & 1.274          & 1.254          & 1.289          & 1.278          \\
Gate                & 1.396          & 1.435          & 1.366          & 1.369          & 1.391          \\
Readout \& Reset    & 1.058          & 1.05           & 1.048          & 1.055          & 1.056          \\ \hline
Combined            & 1.302          & 1.321          & 1.338          & 1.345          & \textbf{1.327} \\ \hline \hline
\end{tabular}
\label{tab:mean_relative_ratio}
\end{table}

Table \ref{tab:mean_relative_ratio} shows that gate errors exhibit the largest impact on logical error rates, with a mean relative ratio of 1.391. In the modified code, gate imperfections in the \(X\)-stabilizer round can propagate through to the \(Z\)-stabilizer round, compounding their effects and significantly increasing logical error rates compared to the original. Combined errors, with a mean relative ratio of 1.327, also show a notable impact, albeit slightly less than gate errors alone, due to the relatively lower influence of readout/reset errors in the mix. Depolarizing errors, with a mean relative ratio of 1.278, affect the system by introducing random flips in qubit states, but their impact is less amplified by the reuse of ancilla qubits. This is because depolarizing errors are inherently probabilistic and do not rely on temporal correlations between rounds. Consequently, the increase in logical error rates due to depolarizing errors is smaller compared to gate and combined errors. Since the ratio is relatively close to 1, the change in logical error rates is not highly significant. Readout/reset errors, with the smallest mean relative ratio of 1.056, have the least impact on logical error rates in the modified code. The ratio being close to 1, indicates that these errors rarely lead to much difference between the original and modified codes. These errors occur during the initialization or measurement of ancilla qubits, and since the ancilla qubits are reset or reinitialized between rounds, their reuse does not significantly amplify the effects of readout/reset errors. In summary, while gate and combined errors dominate in terms of relative impact, the extremely low absolute logical error rates across all error types mean the practical differences between the original and modified codes are low.

\subsubsection{\textbf{Comparing Threshold}}

The primary goal of QECCs is to minimize the logical error rate, even in the presence of significant physical error rates. A key performance metric in this context is the \textit{threshold error rate}, which represents the maximum physical error rate a QECC can tolerate while still reducing the logical error rate. Beyond this threshold, the code's redundancy becomes insufficient to counteract the increasing volume of errors. A higher threshold indicates a more robust QECC, as it signifies the ability to handle greater physical error rates. Thus, comparing the thresholds of the modified surface code with the original approach is critical to assess any changes in performance.

Table~\ref{tab:diff_threshold} presents the threshold error rates for different error types in both the original and modified surface codes. The results indicate minimal changes, with the thresholds for the modified code closely aligning with those of the original. This demonstrates that the modified surface code successfully preserves the fundamental functionality of a traditional surface code. Based on the values in the table, the average difference in threshold between the modified and original codes is approximately $1.86\%$, a negligible variation given the small magnitude of threshold values.

\begin{table}[]
\centering
\caption{Threshold Error Rates for Different Error Types in Original and Modified Surface Codes}
\begin{tabular}{c||cc|c}

\textbf{Error Type} & \textbf{Original} & \textbf{Modified} & \textbf{Diff.\%} \\ \hline \hline
Depolarizing        & 0.085472          & 0.086315          & 0.97              \\ 
Gate                & 0.025381          & 0.026194          & 3.10              \\ 
Readout \& Reset    & 0.092731          & 0.093482          & 0.80              \\ \hline
Combined            & 0.027583          & 0.028319          & 2.59              \\ \hline\hline
\end{tabular}
\label{tab:diff_threshold}
\end{table}

\begin{figure}[h]
    \centering
    \includegraphics[width=1\linewidth]{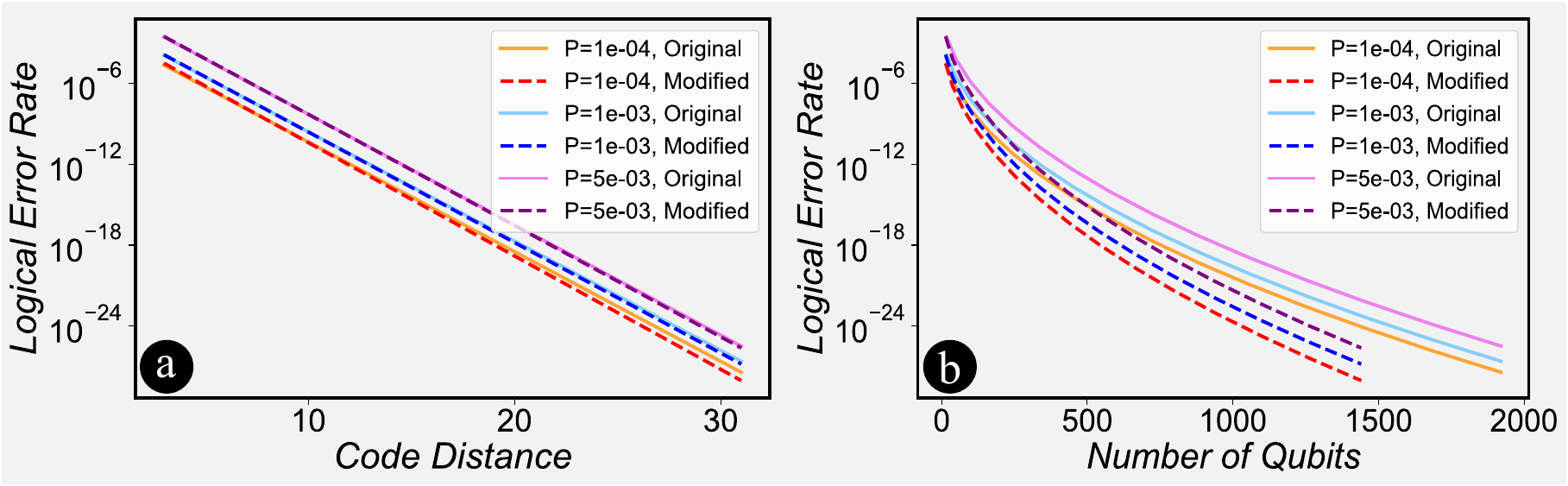}
    \caption{\textbf{Logical Error Rate vs. Code Distance and Number of Qubits for Original and Modified Surface Codes}
    (a) projects how the logical error rate decreases with increasing code distance for both original and modified surface codes under three different physical error rates, considering only the combined error type. The results reveal that higher physical error rates produce higher logical error rates, but these rates decrease consistently with increasing code distance. Both original and modified surface codes exhibit closely aligned trends, with minimal differences, highlighting the comparable effectiveness of the two approaches.
    (b) presents a similar analysis using the number of qubits instead of code distance. It demonstrates that the logical error rate decreases as the number of qubits increases for both original and modified surface codes. However, the modified approach achieves the same logical error rate with fewer qubits than the original approach, underscoring its resource efficiency while maintaining comparable performance.
    }
    \label{fig:logical_vs_distance_and_qubits}
\end{figure}

\subsubsection{\textbf{Performance Analysis with Code Distance}}
Theoretically, as the distance of a surface code increases, the logical error rate is expected to decrease. Consequently, if a user has a target logical error rate in mind, they can select an appropriate surface code distance based on the physical error rate of their system. Figure~\ref{fig:logical_vs_distance_and_qubits} \whitecircled{a}, based on data from previous experiments, projects how the logical error rate decreases with increasing code distance for both original and modified surface codes under three different physical error rates. This analysis considers only the combined error type. The results show that higher physical error rates consistently produce higher logical error rates, but these logical errors decrease as the code distance increases, following the expected trend for both original and modified surface codes. Additionally, the logical error rate trends for the original and modified surface codes are closely aligned, with minimal differences, demonstrating the comparable efficacy of the two approaches.

Figure~\ref{fig:logical_vs_distance_and_qubits} \whitecircled{b} uses the same physical error rates as the previous plot but replaces code distance with the number of qubits. Here, it is observed that as the number of qubits increases, the logical error rate decreases for both original and modified surface codes. However, unlike the previous figure, the modified approach achieves a given logical error rate with fewer qubits compared to the original approach. This indicates that the modified surface code is more resource-efficient while maintaining similar performance.

\section{Conclusion} \label{sec:conclusion}

We introduce qubit overhead reduction in quantum error correcting codes by reusing ancilla qubits for X- and Z-stabilizer measurements. The modified surface code achieves a 25\% reduction in total qubits. 
Furthermore, transitioning from $d$ (original) to $d + 2$ (modified) achieves equivalent or better logical error rates with fewer qubits for $d \geq 13$. These results highlight the modified code's practicality and scalability for resource-efficient quantum systems.

\section*{Acknowledgment}

The work is supported in parts by the National Science Foundation (NSF) (CNS-2129675, CCF-2210963, and DGE-2113839) and gifts from Intel.

\bibliographystyle{IEEEtran}
\bibliography{references}

\end{document}